\documentclass[11pt,onecolumn,draftclsnofoot]{IEEEtran}
\usepackage{amssymb} 

\begin{document}
\input{epsf}

\title{A Global Optimization Approach to Quantum Mechanics}

\author{
\authorblockN{Xiaofei Huang} \\
\authorblockA{School of Information Science and Technology\\
Tsinghua University, Beijing, P.~R.~China, 100084 \\
Email: huangxiaofei@ieee.org}
}

%

\maketitle

\begin{abstract}
This paper presents a global optimization approach to quantum mechanics, 
	which describes the most fundamental dynamics of the universe.
It suggests that the wave-like behavior of (sub)atomic particles could be the critical characteristic
	of a global optimization method deployed by nature
	so that (sub)atomic systems can find their ground states 
	corresponding to the global minimum of some energy function associated with the system.
The classic time-independent Schr\"{o}dinger equation is shown to  be derivable from 
	the global optimization method to support this argument.
\end{abstract}

\section{Introduction}
Could quantum dynamics follow a global optimization process so that
	(sub)atomic systems like molecules can most often evolve into their ground states?
Nature possibly demands such a process so that (sub)atomic systems can find their ground states 
	corresponding to the global minimum of some energy function associated with the system.
If the dynamics of any (sub)atomic system make it evolve into a final state 
	which is the global minimum of its associated energy function,
	then the universe is likely to be deterministic.
Otherwise, if a local minimum is often reached instead, 
	which can be any one of many local minimums of the energy function,
	then it is likely to be probabilistic.
The dynamics which leads to deterministic outcomes for the evolution of (sub)atomic systems
	would make the world more stable.
	

Classic optimization methods such as gradient descent are usually based on local optimization
	to find a local optimum instead of the global one (see Chapter 5 in \cite{Pardalos02} for a comprehensive survey of the classic methods for nonlinear programming).
The fundamental principle of those methods is based on the iterative improvement of solutions 
	which suffers from the local optimum problem.
The key characteristic of their operations is to assign each variable a precise value at any given time.
If nature deploys a local optimization process to define the dynamics of atomic systems 
	where each object (particle) in the system has a precise location (or momentum) at any given time,
	then molecules including proteins will unlikely be stable 
	because their final states could be probabilistic and unpredictable.
The system can be trapped into one local minimum or another, sensitive to initial conditions and perturbations.

Cooperative optimization is a general optimization method for finding global optimal solutions instead of local ones~\cite{Huang03-SIGACT,Huang03Greece,HuangBookCCO,HuangDAGM04,HuangCP04}.
It does not struggle with local minima and offers us a complete departure from conventional local optimization methods.
It was originally developed for solving real-world, NP-hard optimization problems raised from
	the areas such as communications~\cite{HuangISIT05}, computer vision~\cite{Huang03Greece,HuangDAGM04}, and image processing~\cite{Huang04ICIP}.
In real-world applications, 
	cooperative optimization is remarkably successful and has often outperformed state-of-the-art algorithms.
The key characteristic of its operations is to make soft decisions at assigning variable values at any given time;
	namely, the preference of picking each value for a variable is weighted numerically.
This paper demonstrates that one of the most fundamental equations in physics, 
	the time-independent Schr\"{o}dinger equation~\cite{Schrodinger26a,Schrodinger26b,Schrodinger26c},
	can be derived from the dynamic equations of cooperative optimization.
Such a connection between the quantum dynamic equation and a global optimization algorithm 
	offers us new insight into quantum mechanics.

We postulate that quantum dynamics follows a global optimization process 
	such that (sub)atomic systems can most often find their ground states.
The wave-like behavior of (sub)atomic particles, besides Max Born's probabilistic interpretation~\cite{Born26}, 
	could also be the critical characteristic
	of the global optimization process making soft decisions at assigning variable values at any given time
	so that the global minimum instead of a local minimum can be found.
	
\section{Cooperative Optimization For Finding Global Optima}

Cooperative optimization is a general principle for finding global optima of a multivariate function $E(x_1, x_2, \ldots, x_n)$.
It utilizes a function in simple forms, such as $\psi(x_1, x_2, \ldots, x_n) = \sum_i \Psi(x_i)$, 
	and iteratively refines the function $\Psi(x)$ as the lower bound of the multivariate function $E(x)$.
At given time $t$, if the lower bound function $\Psi(x, t)$ has been tightened enough 
	so that its global minimum equals to the global optimum of $E(x)$,
	i.e.,
\[ \min_{x} \Psi(x, t) = \min_{x} E(x) \ , \]
	then the global minimum of $E(x)$
	is found which is the same as the global minimum of $\Psi(x, t)$, 
\[ \arg \min_{x} E(x) = \arg \min_{x} \Psi(x, t) \quad \mbox{ if }\min_{x} \Psi(x, t) = \min_{x} E(x) \ . \]
The global minimum of $\Psi(x, t)$ of the form $\sum_i \Psi(x_i,t)$ can be easily computed as
\[ x^{*}_i(\Psi(x, t)) = \arg \min_{x_i} \Psi_i (x_i, t) \quad \mbox{for $i = 1, 2, \ldots, n$} \ . \]

Assume that the multivariate function $E(x)$, often referred to as the energy function in physics,
	can be decomposed as the aggregation of a number of sub-energy functions,
\[ E(x) = \sum_i E_i(x) \ . \]
Assume further that $\Psi(x, t)$ of the form $\sum_i \Psi(x_i, t)$ is a lower bound function of $E(x)$, $\Psi(x, t) \le E(x)$.
Let $\Psi(x_i, t+1)$, for $i=1,2,\ldots, n$, be computed as follows
\begin{equation}
\Psi_i (x_i, t+1) = \min_{x_j \in  X_i \setminus{x_i}}\left( \left(1 - \lambda \right) E_i (x)  +
\lambda \sum_{j} w_{ij} \Psi_j(x_j, t)\right) \ .
\label{cooperative_optimization}
\end{equation}
Then the new function $\Psi(x, t+1) = \sum_i \Psi_i(x_i, t+1)$ is also a lower bound function of $E(x)$, guaranteed by theory~\cite{HuangBookCCO}.
In the above equation, $\lambda$ and $w_{ij}$ ($1 \le i,j \le n$)
	are coefficients of the linear combination of $E_i$ and $\Psi_j (x_j, t)$.
$\lambda$ satisfies $0 \le \lambda < 1$ and $w_{ij}$s satisfy $w_{ij} \ge 0$ and $\sum_i w_{ij} = 1$ for $1 \le i, j \le n$.

The difference equations~(\ref{cooperative_optimization}) define
	the dynamics of cooperative optimization.
The original minimization problem $\min_x E(x)$ has been divided into $n$ sub-problems of minimization (see (\ref{cooperative_optimization})).
Those sub-problems can be solved in parallel in implementation.
The function $\Psi_i (x_i, t)$ is the solution at solving the $i$th sub-problem.
The energy function of the $i$th sub-problem, denoted as $\tilde{E}_i(x)$,
	is a linear combination of the original sub-energy $E_i(x)$ and the solutions from solving sub-problems other than the $i$th,
	i.e.,
\[ \tilde{E}_i(x) = \left(1 - \lambda \right) E_i (x)  + \lambda \sum_{j} w_{ij} \Psi_j(x_j, t)	\ . \]
The cooperation among solving those sub-problems of minimization is thus achieved
	by having each sub-problem compromising its solution with the solutions from others.
$\tilde{E}_i(x)$ is called the modified objective function for the sub-problem $i$.

The coefficient $\lambda$ is a parameter for controlling the cooperation at solving the sub-problems and is called the cooperation strength.
A high cooperation strength leads to strong cooperation at solving the sub-problems
	while a lower cooperation strength leads to weak cooperation.
The coefficients $w_{ij}$ control the propagation of the sub-problem solutions $\Psi_i$ 
	in the modified objective functions $\tilde{E}_i(x)$ (details in \cite{HuangBookCCO}).

The function $\Psi_i (x_i, t)$ can be understood as the soft decision at assigning the variable $x_i$
	at minimizing $\tilde{E}_i(x)$.
The most preferable value for variable $x_i$ at time $t$ is $\arg \min_{x_i} \Psi_i (x_i, t)$.

The cooperative optimization theory~\cite{HuangBookCCO} guarantees that 
	the dynamic system defined by the difference equations~(\ref{cooperative_optimization}) 
	has a unique equilibrium and converges to it with an exponential rate 
	regardless of initial conditions and is insensitive to perturbations.

Without loss of generality,  assume that all energy functions including the sub-energy functions $E_i(x)$ are nonnegative functions.
Then the cooperative optimization theory tells us that the lower bound function $\Psi$ 
	computed by (\ref{cooperative_optimization}) can be progressively tightened, 
\[ \Psi(x, t=0) \le \Psi(x, t=1) \le \Psi(x, t=2) \le \ldots \le \Psi(x, t=k) \le E(x) \ , \]
	when we choose the initial condition as $\Psi_i(x_i, t = 0) = 0$, for $i=1,2,\ldots, n$.
	
With certain settings of the cooperation strength $\lambda$ and the solution propagation coefficients $w_{ij}$,
	the time-independent Schr\"{o}dinger equation can be derived in mathematical form 
	from the difference equation~(\ref{cooperative_optimization}) of cooperative optimization.
The following two sections offer the detail of the settings and the derivation.

\section{Variations of Cooperative Optimization}

If the energy function $E(x_1, x_2, \ldots, x_n)$ is of the following form
\begin{equation}
E(x_1,x_2,\ldots,x_n)=\sum^{n}_{i=1} e_i (x_{i}) + \sum^{n}_{i=1} \sum^{n}_{j=1, j \not = i} e_{ij} (x_i, x_j),
\label{binary_cost_function1}
\end{equation}

Let the decomposition of $E(x)$ be
\begin{equation}
E_i = e_i (x_i) + \sum_{j,~j \not = i} e_{ij} (x_i, x_j), \quad \mbox{ for } i = 1, 2,\ldots, n
\label{sub_binary_cost_function1}
\end{equation}

The difference equations~(\ref{cooperative_optimization}) become
\begin{equation}
\Psi_i(x_i, t+1)= (1-\lambda) e_i(x_i)+\lambda w_{ii}\Psi_i(x_i, t) + \sum_{j, j \not = i} \min_{x_j} 
   \left( (1-\lambda) e_{ij}(x_i,x_j) + \lambda w_{ij}\Psi_j(x_j, t)  \right) \ .
\label{diff_equations1}
\end{equation}

Choosing the coefficients $w_{ij}$ as $w_{ii} = 0$ and $w_{ij} =a$ (a positive constant), for $i \not = j$,
	we have
\begin{equation}
\Psi^{'}_i(x_i, t+1)= e_i(x_i)+ \sum_{j, j \not = i} \min_{x_j} 
   \left( e_{ij}(x_i,x_j) + \alpha \Psi^{'}_j(x_j, t)  \right) \ ,
\label{diff_equations1a}
\end{equation}
where $\alpha = \lambda a$ and $\Psi^{'}_i(x_i, t) = \Psi_i(x_i, t) / (1 - \lambda)$.
In (\ref{diff_equations1a}), the parameter $\alpha$ controls the cooperation strength.

To possibly improve the convergence of the cooperative optimization defined by the difference equations~(\ref{diff_equations1a}),
	we can offset both sides of the equation by a value, denoted as $z_i(t+1)$, for $i=1,2,\ldots, n$.
One possible choice for $z_i(t+1)$ is 
\[ z_i(t+1) = \min_{x_i} \Psi^{'}_i(x_i, t+1) \ . \]

Let $\Psi^{''}_i(x_i, t+1) = \Psi^{'}_i(x_i, t+1) - z_i(t+1)$, 
	the difference equation~(\ref{diff_equations1a}) becomes
\begin{equation}
\Psi^{''}_i(x_i, t+1)= e_i(x_i)+ \sum_{j, j \not = i} \min_{x_j} 
   \left( e_{ij}(x_i,x_j) + \alpha \Psi^{''}_j(x_j, t) \right) - z_i(t+1) \ .
\label{diff_equations2}
\end{equation}

Dividing the both sides of the above equation by a negative value $- \hbar$ ($\hbar > 0$)
	followed by taking the exponent of the both sides, we have
\begin{equation}
\psi_i(x_i, t+1)= \frac{1}{e^{z_i(t+1)/\hbar}} e^{-e_i(x_i)/\hbar} \prod_{j, j \not = i} 
	\left(\max_{x_j} e^{-e_{ij}(x_i,x_j)/\hbar} \psi_j(x_j, t)^{\alpha}\right) \ ,
\label{diff_equations_max_product_b}
\end{equation}
where 
\[ \psi_i(x_i, t+1) = e^{-\Psi^{''}_i(x_i, t+1)/\hbar}. \] 
The function $\psi_i(x_i, t)$ is the soft decision at assigning the variable $x_i$.
It is called the soft assignment function in this paper.
For each value of $x_i$, the function $\psi_i(x_i, t)$ measures the preference 
	of assigning that value at the iteration time $t$ for that variable.
A high function value represents a strong preference of assigning that value for the variable.

Using the approximation 
\[ e^{-\max_{x} g(x)/\hbar} \approx \sum_{x} e^{-g(x)/\hbar} \ , \]
which has the property of
\[ \lim_{\hbar \rightarrow 0} e^{-\max_{x} g(x)/\hbar}/ \sum_{x} e^{-g(x)/\hbar} = 1 \ , \]
	and choosing the parameter $\alpha = 2$,
	the difference equations~(\ref{diff_equations_max_product_b}) can be rewritten as
\begin{equation}
\psi_i(x_i,t+1)= \frac{1}{Z_i(t+1)} e^{-e_i(x_i)/\hbar} \prod_{j, j \not = i} 
	\left(\sum_{x_j} e^{-e_{ij}(x_i,x_j)/\hbar} |\psi_j(x_j, t)|^2\right) \ ,
\label{diff_equations_max_product_c}
\end{equation}
where $Z_i(t+1) = e^{z_i(t+1)/\hbar}$.

If $\psi_j(x_j, t) \in \mathbb{C}$ (the complex domain), 
	the difference equation~(\ref{diff_equations_max_product_c}) still works for global optimization.
In this case, $|\psi_j(x_j, t)|^2 = \psi^{*}_j(x_j, t)\psi_j(x_j, t)$ represents the soft decision at assigning the variable $x_j$.
The best candidate value for assigning $x_j$ at time $t$ is the one of the highest function value $|\psi_j(x_j, t)|^2$.
Any variable value other than the best one may also have a positive function value $|\psi_j(x_j, t)|^2$
	representing the degree of preference of assigning that value to the variable $x_j$.
If only one value of $x_j$, say $x^{*}_j$, has a positive function value $|\psi_j(x^{*}_j, t)|^2 > 0$
	while all other variable values have $|\psi_j(x_j, t)|^2 = 0$ ($ x_j \not = x^{*}_j$),
	then the decision at assigning $x_j$ becomes a hard one instead of a soft one.

As discussed above, given any $i$ and $t+1$, $z_i(t+1)$ can be any value that can possibly improve the convergence.
One way to do that is to choose a value for $Z_i(t)$, which is equivalent to choose a value for $z_i(t)$, such that
	$\sum_{x_i} |\psi_i(x_i, t)|^2 $ is bound to a constant, say $1$.
To be more specific, we choose $Z_i(t)$ as
\[ Z_i(t) = \sum_{x_i} |\psi_i(x_i, t)|^2 \ . \]
With such a choice, 
\[ \sum_{x_i} |\psi_i(x_i, t)|^2 = 1 \ . \]
Hence, this choice makes $|\psi^{(k)}_i(x_i,t)|$, for $i=1,2,\ldots, n$, 
	have the property of a probability density function.
$Z_i(t+1)$ is thus called the normalization factor.
(The specific choice of $Z_i(t+1)$ makes no difference at the optimization power of the cooperative optimization.)

If all variables $x_i$s are in a continuous domain, the difference equations~(\ref{diff_equations_max_product_c}) become
\begin{equation}
\psi_i(x_i, t+1) = \frac{1}{Z_i(t+1)} e^{-e_i(x_i) /\hbar} \prod_{j, j \not = i} 
\int d x_j~e^{-e_{ij}(x_i, x_j) /\hbar} |\psi_j (x_j, t)|^2 \ . 
\label{update_rule4}
\end{equation}

\section{A Continuous Time Version of the Cooperative Optimization algorithm}

Let the soft assignment function of $x_i$ at time $t$ be $\psi_i (x_i, t)$.
Let $\Delta t$ be an infinitesimal positive value and the soft assignment function at $t + \Delta t$  be $\psi_i (x_i, t + \Delta t)$.
The difference equations~(\ref{update_rule4}) of the cooperative optimization algorithm in a continuous time version become
\begin{equation}
\psi_i(x_i, t + \Delta t) = \frac{1}{Z_i (t + \Delta t)} \psi_i(x_i, t) e^{-(\Delta t/\hbar)e_i(x_i)} \prod_{j, j \not = i} 
\int d x_j~e^{-(\Delta t/\hbar)e_{ij}(x_i, x_j)} |\psi_j (x_j, t)|^2 \ . 
\label{update_rule5}
\end{equation}
From (\ref{update_rule5}) we have
\[ \lim_{\Delta t \rightarrow 0} \psi_i(x_i, t + \Delta t) = \psi_i(x_i, t) \ . \]

A dynamic system described by (\ref{update_rule5}) is a dissipative system, not a conservative system.
It will evolve toward its equilibriums over time.
It will be shown in the following that 
	the dynamic equation~(\ref{update_rule5}) at its equilibrium is, in fact, the time-independent Schr\"{o}dinger equation.

Assume that the initial condition is $\psi_i (x_i, t_0) = \delta (x_i - a_i)$, for $i = 1,2,\ldots, n$,
	where $\delta (x_i - a_i)$ is the delta function defined as
\[ \delta (x_i - a_i) = 0, \mbox{ when } x_i \not = a_i, \quad \mbox{ and } \int \delta (x_i - a_i)~d x_i = 1 \ . \]
and $a_i$ is not the optimal value, $a_i \not = x^{*}_i$.
Then $(\psi_i (x_i, t_0))$ is a undesired stationary state of the difference equation system~(\ref{update_rule5}).
To improve the performance of the system,
	we spread the soft assignment function by a smoothing kernel $K(x)$, i.e.,
\[ \int K(u - x_i) \psi_i (u) ~d u \Rightarrow \psi_i(x_i) \ . \]

If $x_i$ is in the one dimensional space $\mathbb{R}$, 
	we can choose the following Gaussian function as $K(x)$
\begin{equation}
K(x) = \frac{1}{\sqrt{2 \pi \Delta t} \sigma_i} e^{ - x^2 / 2 \sigma^2_i \Delta t} \ . 
\label{gaussian_kernel_1d}
\end{equation}

With the soft assignment function spreading, the dynamic equation~(\ref{update_rule5}) becomes
\begin{eqnarray}
\psi_i(x_i, t + \Delta t) &=& \frac{1}{Z_i (t + \Delta t)} \int d u~\frac{1}{\sqrt{2 \pi \Delta t} \sigma_i} e^{ - (u - x_i)^2 / 2 \sigma^2_i \Delta t} \psi_i(u, t) e^{-(\Delta t /\hbar) e_i(u)} \nonumber \\
	 && \times \prod_{j, j \not = i} \int d x_j~e^{-(\Delta t /\hbar)e_{ij}(u, x_j)} |\psi_j (x_j, t)|^2  \ , 
\label{update_rule10}
\end{eqnarray}
Expanding the right side of the above equation into a Taylor series with respect to $\Delta t$ and let $\Delta t \rightarrow 0$,
	we have the new differential equations for the cooperative optimization algorithm in a continuous time version,
\begin{equation}
\frac{\partial \psi_i (x, t)}{\partial t}
	=  \frac{\sigma^2_i}{2}\frac{\partial^2 \psi (x_i, t)}{\partial x^2_i} - V_i (x_i)\frac{1}{\hbar} \psi_i(x_i, t) + \varepsilon_i (t) \psi_i(x_i, t), \quad \mbox{ for each $i$} \ , 
\label{update_rule11a}
\end{equation}
where 
\[ V_i(x_i) = e_i(x_i) + \sum_{j, j \not = i} 
  \int d x_j~e_{ij}(x_i, x_j) |\psi_j (x_j, t)|^2 \quad \mbox{and} \quad \varepsilon_i (t) = -\frac{d~Z_i(t)/d~t}{Z^2_i(t)} \ . \]

Let the operator $\nabla^2_i$ be defined as
\[ \nabla^2_i \psi (x, t) = \frac{\partial^2 \psi (x_i, t)}{\partial x^2_i} \ , \]
and $H_i$ be an operator on $\psi (x_i, t)$ defined as
\begin{equation}
H_i = -\frac{\hbar \sigma^2_i}{2} \nabla^2_i +  V_i(x_i) \ . 
\label{Hamiltonian}
\end{equation}
Then the equation~(\ref{update_rule11a}) can be rewritten as
\begin{equation}
\frac{\partial \psi_i (x, t)}{\partial t}
	= - \frac{1}{\hbar} H_i \psi_i(x_i, t) + \varepsilon_i (t) \psi_i(x_i, t)   \ . 
\label{update_rule11}
\end{equation}

When the differential equations~(\ref{update_rule11}) evolve into a stationary state (equilibrium), 
	the differential equations~(\ref{update_rule11}) become
\begin{equation}
E_i \psi_i(x_i, t)  =  H_i \psi_i(x_i, t) , \quad \mbox{ for $i=1, 2, \ldots, n$} \ , 
\label{update_rule12}
\end{equation}
where $E_i$, $E_i = \hbar \varepsilon_i$, is a scalar.

For a physical system consisting of $n$ particles, 
	let $x_i$ be the position of particle $i$, $1 \le i \le n$, in the one dimensional space $\mathbb{R}$.
Let
\[ \sigma^2_i = \hbar / m_i, \]
where $m_i$ is the mass of particle $i$.
Then equations~(\ref{update_rule12}) become
\begin{equation}
E_i \psi_i(x_i, t)  =  \left( -\frac{\hbar^2}{2 m_i} \nabla^2_i + V_i(x_i) \right) \psi_i(x_i, t) , \quad \mbox{ for $i=1, 2, \ldots, n$} \ . 
\label{stationary_state}
\end{equation}
They are the condition for the physical system being in a stationary state when its dynamics is defined by the cooperative optimization.
Equation~(\ref{stationary_state}) is also the time-independent Schr\"{o}dinger equation.
(It is straightforward to generalize this derivation to three dimensions, but it does not yield any deeper understanding.)
From the equation~(\ref{stationary_state}) we can see that 
	the soft assignment function $\psi_i(x_i, t)$  is the wavefunction in the Schr\"{o}dinger equation.

If we choose more advanced forms for $\Psi(x)$ where two (or more) variables, say $x_i$ and $x_j$, are clustered,
	together with their soft assignment functions $\Psi_i(x_i)$ and $\Psi_j(x_j)$, i.e.,
	$\Psi_i(x_i), \Psi_j(x_j) \Rightarrow \Psi_{ij}(x_i, x_j)$,
	then better lower bound functions in terms of tightness can be found  
	and there is a higher chance for the cooperative optimization algorithm to find a global minimum~\cite{Huang03Greece}.
However, more variable clustering leads to higher computational cost as a tradeoff.
The quantum entanglement phenomena could possibly be understood as variable clustering 
	in cooperative optimization to increase the chance of finding the global optimum.

\section{Conclusions and Future Research}

This paper derived the time-independent Schr\"{o}dinger equation being
	as the dynamic equation of cooperative optimization at its equilibrium in a continuous-time version for continuous variables.
Cooperative optimization is a general method for finding global optima with
	the key characteristic of soft decision making in assigning variables.
The soft assignment functions used by cooperative optimization 
	can be interpreted as the wavefunctions in the Schr\"{o}dinger equation.
It could be a critical feature for (sub)atomic systems to have deterministic final states, namely their ground states.

In classic quantum mechanics, wavefunction spreading is understood 
	to have the purpose of preventing electrons from collapsing into the nucleus caused by the Coulomb force.
In the global optimization approach, it also serves the purpose of improving convergence property 
	of quantum dynamics at finding global optima.

For a closed system, we can use the time dependent Schr\"{o}dinger equation to describe its evolution (nonrelativistic).
Usually, a closed system is a conservative system.
However, it is hard to find a closed system in nature because the (sub)atomic particles in the system
	will inevitably interact with their environment and the system becomes dissipative.
Could global optimization be an emerging property of such a dissipative process
	so that (sub)atomic systems can most often find their ground states?
Could the wave-like property of (sub)atomic particles play a key role at defining such a global optimization process?
Could the dynamic equation offered by cooperative optimization be suitable for describing the global optimization process at a proper level?
These are interesting questions worth pursuing with future research.

\nocite{Messiah99,Tegmark01,Seife05}


\end{document}